\documentclass{article}
\usepackage{spconf,amsmath,graphicx}

\usepackage[utf8]{inputenc}
\usepackage{siunitx}
\usepackage{multirow}

\newcommand{\bs}[1]{\boldsymbol{#1}}

\title{Full-Reference Speech Quality Estimation with Attentional Siamese Neural Networks}
\name{Gabriel Mittag$^1$, Sebastian Möller$^{1,2}$\thanks{The work on this paper was largely supported by the BMBF, Grant 01IS17052.}}
\address{$^1$ Quality and Usability Lab, Technische Universität Berlin, Berlin, Germany\\
$^2$ Language Technology, Deutsches Forschungszentrum für Künstliche Intelligenz (DFKI), Berlin, Germany}

\begin{document}
%
\maketitle
\begin{abstract}
In this paper, we present a full-reference speech quality prediction model with a deep learning approach. The model determines a feature representation of the reference and the degraded signal through a siamese recurrent convolutional network that shares the weights for both signals as input. The resulting features are then used to align the signals with an attention mechanism and are finally combined to estimate the overall speech quality. The proposed network architecture represents a simple solution for the time-alignment problem that occurs for speech signals transmitted through Voice-Over-IP networks and shows how the clean reference signal can be incorporated into speech quality models that are based on end-to-end trained neural networks.
\end{abstract}
%
\section{Introduction}
\label{sec:intro}
The perceived quality of transmitted speech is traditionally evaluated in auditory listening experiments, where speech samples are rated by naïve test participants on a 5-point rating scale. The average across all test participants then yields the so-called mean opinion score (MOS). Because these experiments are time-consuming and costly, instrumental measurements have been developed that estimate an objective MOS. These models are used to monitor communication networks, by sending a test speech signal through the channel and predicting the speech quality of the recorded output signal. The current recommendation for objective speech quality assessment by the International Telecommunication Union (ITU-T) is P.863 (POLQA) \cite{ITUTRec.P863}, which is the successor of the widely used PESQ model. POLQA consists of a temporal alignment model that determines the delay between the clean reference and the degraded output signal and a perceptual model that predicts the speech quality. The perceptual model compares the aligned reference with the degraded signal by transforming both signals to an internal representation that takes into account the pitch and loudness perception of the human auditory system.

Recently, several speech quality models based on deep neural networks have been presented. These networks learn suitable features automatically, without the need for manually engineered indicators. In \cite{Soni2016} a model was presented that uses learned features from a deep autoencoder to predict speech quality. In \cite{Fu2018} a model was proposed that is based on an LSTM network (Long Short Term Memory) that predicts the quality on a frame-level. In \cite{Ooster2018, Ooster2019} the output of a deep automatic speech recognizer is used to estimate speech quality. In \cite{Avila2019} three different neural network approaches are proposed, where a deep fully connected network obtained the best results. In \cite{Cauchi2019} a model that combines modulation energy features with an LSTM network is proposed. In \cite{Shan2019} a deep belief network is trained to produce a pseudo-reference, which is then used to calculate distance features to predict speech quality. In \cite{Catellier2019} a network based on a 1-D CNN layer (convolutional neural network) is presented, and in \cite{mittag2019ic} we presented a model that firstly uses a CNN network to estimate the quality on a frame-level, followed by an LSTM network that estimates the overall quality. 

These models are all non-intrusive, which means that they only rely on the degraded output signal without the need for a clean reference. In this paper, we investigate deep learning architectures that incorporate the clean reference into the network to improve the prediction accuracy and compare it to the current state-of-the-art model POLQA. One of the main challenges in full-reference speech quality prediction is the alignment of both signals. When a speech signal is sent through a communication network it is exposed to a channel latency. In VoIP (Voice over IP) networks the jitter buffer tries to compensate for delayed packets in the network by stretching and shrinking the speech signals in time. The jitter buffer can also drop packages when the delay reaches a certain threshold, to reduce the delay. This introduces a variable delay across the speech signal that is difficult to determine. POLQA uses a complex time alignment algorithm \cite{beerends2013perceptual} that consists of five major blocks: filtering, pre-alignment, coarse alignment, fine alignment, and section combination. In this paper, we present an end-to-end trained Siamese neural network that uses an attention mechanism to align both signals.\footnote{Open-sourced at www.github.com/gabrielmittag/NISQA}

\section{Model}
\label{sec:model}
\begin{figure}[!ht]
    \begin{center}
        \includegraphics[width=\linewidth]{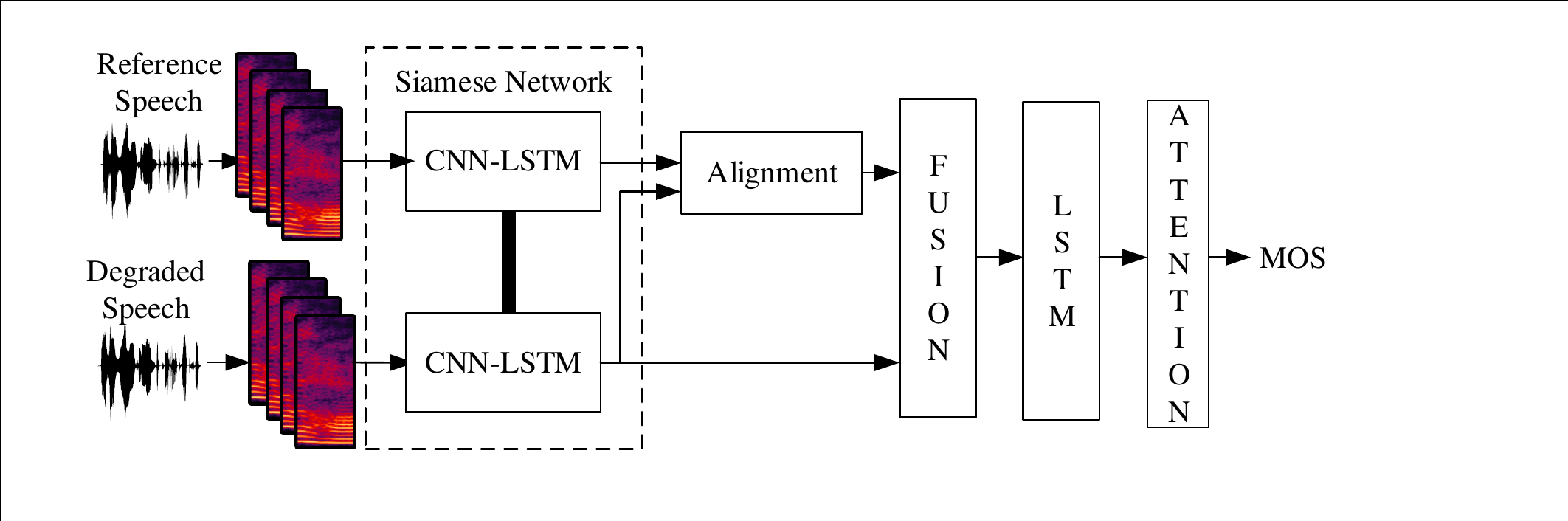}
    \end{center}
    \caption{Block diagram of the proposed full-reference model}
    \label{fig:blockdia}
\end{figure}
The model firstly calculates a feature representation of the speech signals through a CNN-LSTM network. Both, the clean reference and the degraded signal, are sent through this network in parallel. The network shares the weights and thus optimizes them for both inputs. This type of architecture is also called Siamese neural network and allows for calculating features that are comparable for different input vectors. They have, for example, proven to be useful for image quality prediction \cite{bosse} and for the alignment of speaker verification tasks \cite{verif}. The calculated features are then used to align the reference to the degraded signal. The features of both signals are combined in a fusion layer and send through another LSTM network to estimate the overall quality. Additionally to the LSTM outputs of the last time step we also applied a further attention layer that is supposed to draw the network's attention to distorted segments of the speech signal. The LSTM output of the last time step is then combined with a weighted sum of the outputs of all time steps to estimate the overall speech quality. An overview of the model can be seen in Figure \ref{fig:blockdia}.
\subsection{Siamese CNN-LSTM network}
At first log-mel spectrograms are calculated from the speech signals. Then 150 ms long speech segments with the dimensions 48x15 and a hop size of 10 ms are extracted. These segments are then used as input for the CNN network. The CNN consists of six convolutional layers, followed by batch normalization, ReLU, and max-pooling layers, and is identical to the CNN presented in \cite{mittag2019c}. The last layer is a fully-connected layer with an output of size 20, which is used as input for the subsequent LSTM network. We apply an LSTM layer to learn time dependencies between consecutive frames and in the hope to give the model some contextual awareness. This bidirectional LSTM consists of one layer with 20 hidden units in each direction, resulting in an output size of 40. Because the CNN-LSTM network calculates the same features for the reference and the degraded signal and is trained to estimate the overall quality, it should, in theory, result in features that are optimized to find perceptual differences between both signals.

\subsection{Alignment with attention}
The idea for the alignment is that reference features of a certain time step will be similar to the corresponding time step features of the degraded signal. To determine the similarity between different time scales many different score functions have been proposed. A basic one that is often used with siamese networks is the L1-norm of the difference between the features. For a certain time step of the degraded signal it can be described as
\begin{equation}
    \bs{d}_i = ||\bs{s} - \bs{h}_i||, 
\end{equation}
where $\bs{s}$ is the LSTM output of the degraded signal at a fixed time step, $\bs{h}_i$ is the output of the reference signal at time step $i$, and $\bs{d}_i$ are the L1-norm differences at time step $i$ of the reference signal. Since we have $H=40$ output features of the LSTM network we calculate the average L1-norm across all feature values, resulting in the alignment score $e_i$:
\begin{equation}
    e_i = - \frac{1}{H} \sum^H_{j=1} \bs{d}_i(j).
\end{equation}
The L1-norm will be small for a large similarity, because of this we negate the results to obtain an alignment score that is high for similar features.

Another basic score function, which is often used in attention layers for natural language processing tasks is the dot product:
\begin{equation}
    e_i = \bs{s}^T \bs{h}_i. 
\end{equation}
The resulting alignment score can be seen as a correlation or similarity measure between features of the degraded and the reference signal. 

We calculate the alignment scores $\bs{e}$ for each possible combination of time steps, resulting in a $N\times M$ matrix, where $N$ is the number of time steps of the degraded signal and $M$ the number of time steps of the reference signal. Then for each time step of the degraded signal, we find the index in the reference signal with the highest alignment score $e_i$ and save the feature values at this index as a new aligned reference feature vector. This approach can be seen as a hard-attention mechanism, where the network only focuses on the reference time step with the highest similarity. 
\subsection{Feature fusion}
The CNN-LSTM output features of the degraded signal are then combined with the aligned output features of the reference signal. In \cite{Cauchi2019} it was shown that - although the network should be able to find this automatically - it can be helpful for the regression task to explicitly include the difference between the features. Because of this we concatenate the features as follows: $f =$ \texttt{concat}($f_\mathrm{deg}$, $f_\mathrm{ref}$, $f_\mathrm{deg}-f_\mathrm{ref}$), which yields a fused feature vector $f$ with output size $3\cdot H=120$ and length $N$.
\subsection{Time pooling}
The fused features are then used as input for a second bidirectional LSTM network with one layer and 2x256 hidden units that estimates the overall speech quality. Since the perceived quality of a speech signal can vary over time, we add another attention layer after the second LSTM network. The attention layer should find the time steps that especially stand out, for example short-time distortions. The assumption is that test participants in a listening test also put their attention to these segments of the speech signal and give them more weight than other parts of the signal (e.g. silent segments). As score we use a linear attention function:
\begin{equation}
    k_i = \bs{w^T} \bs{g}_i + b, 
\end{equation}
where $\bs{g}_i$ is the output of the second LSTM network at time step $i$, $\bs{w}$ is a vector of size 512 and $b$ is a scalar, both are learned during training. The linear function gives a scalar weight $k_i$ for each time step of the degraded speech signal. This weight is then normalized by a softmax layer as follows:
\begin{equation}
    a_i = \frac{\mathrm{exp}(k_i)}{\sum_{j=1}^N \mathrm{exp}(k_j)}.
\end{equation}
The weighted feature vector is then formed as $\bs{g}_w = \sum^{N}_{i=1} a_i \bs{g}_i$ and concatenated with the features of the last time step $g_N$ ($g_a =$ \texttt{concat}($g_N$, $g_w$)). $g_a$ is then used as input for the final fully connected layer that estimates the speech quality. Additionally we also use a simple mean over all time steps to compare the attention layer to a baseline.
%
%
\section{Databases}
\label{sec:db}
Overall 21 super-wideband databases were available. They have been rated in auditory listening tests according to ITU-T P.800 \cite{ITUTRec.P.800} and contain 6-12s long speech signals. 17 of the databases were part of the ITU-T P.OLQA competition and contain a wide variety of different degradations, including live recordings. We further conducted P.800 listening tests for four additional databases that also contain simulated and live conditions. 11 of the databases were used for training (all from the POLQA pool, DBs: 101, 201, 202, 401, 301, 302, 501 502, 601, 602, and EXP4 / 3544 files) and 10 for validation (see Table \ref{tab:results_2} / 3436 files). 

Since deep learning models need a lot of data to be trained, we generated an additional pretraining database for which the MOS values were predicted with POLQA. The pretraining database contains overall 100,000 speech samples generated from 5,000 reference files. The simulated distortions contain different codecs (G711, G722, AMR-NB, AMR-WB, Opus, EVS) with different bitrates, background noises, white noise, amplitude clipping, time clipping, packet-loss, and combinations of these distortions. 
%

\section{Experiments and Discussion}
\label{sec:exp}

\begin{table}[htb]
\caption{Prediction error of the proposed models across all validation databases. \textbf{LM}: L1-norm with mean pooling, \textbf{LL}: L1-norm with linear attention, \textbf{DM}: Dot with mean pooling, \textbf{DL}: Dot with linear attention, \textbf{SM}: Single-ended with mean pooling, \textbf{SL}: Single-ended with linear attention}
\vspace{0.2cm}
\label{tab:results_1}
\centering
\small
\begin{tabular}{l|llllll}
              & LM            & LL   & DM   & DL   & SM   & SL   \\ \hline
Average RMSE* & \textbf{0.18} & 0.21 & 0.20 & 0.20 & 0.22 & 0.22 \\
Max RMSE*     & \textbf{0.29} & 0.33 & 0.31 & 0.30 & 0.32 & 0.32
\end{tabular}
\end{table}

After the training with the pretraining database, which only contains objective MOS, the model is trained with the subjective MOS values of the 11 training databases\footnote{The pretraining increased the model accuracy significantly. An analysis is however not included due to the lack of space.}. The results are evaluated in terms of the \textit{epsilon-insensitive RMSE} RMSE* after a third-order polynomial monotonous mapping, according to ITU-T Rec. P.1401 \cite{ITUTRec.P.1401}. The calculation of the RMSE* is similar to the traditional \textit{root mean square error} (RMSE) but takes into account the confidence interval of the subjective MOS scores (see P.1401 eq. (7.29)). The mapping compensates for offsets, different biases, and other shifts between scores from the individual experiments, without changing the rank order.

\begin{table*}[htb]
\caption{Validation results with L1-norm alignment and mean time pooling, compared to P.OLQA and VISQOL}
\vspace{0.2cm}

\centering
\label{tab:results_2}

\resizebox{0.75\textwidth}{!}{%
\begin{tabular}{{lcccc|ccc|ccc|ccc}}
 & 
 &
 &
Files &
Listeners &
\multicolumn{3}{c|}{Proposed (LM)} &
\multicolumn{3}{c|}{POLQA} &
\multicolumn{3}{c}{VISQOL}  \\

Validation Databases & 
Lang &
Con &
Per Con &
Per File &
{$r_\mathrm{pearson}$} & 
\small{RMSE} & 
\small{RMSE*} & 
{$r_\mathrm{pearson}$} & 
\small{RMSE} & 
\small{RMSE*} & 
{$r_\mathrm{pearson}$} &                                                                                                                                            
\small{RMSE} &                                                                                                                                            
\small{RMSE*} \\ \hline            
                                                                                                                   
103\_ERICSSON                              &  sv & 54  & 12 & 8        & 0.88    & 0.46   & 0.24  & 0.90   & 0.42  & 0.24     & 0.37  & 0.81   & 0.55    \\
203\_FT\_DT                                &  fr & 54  & 4  & 24       & 0.91    & 0.44   & 0.29  & 0.86   & 0.49  & 0.29     & 0.42  & 0.89   & 0.73    \\
303\_OPTICOM                               &  en & 54  & 4  & 24       & 0.92    & 0.45   & 0.19  & 0.92   & 0.45  & 0.16     & 0.35  & 1.46   & 0.67    \\
403\_PSYTECHNICS                           &  en & 48  & 24 & 8        & 0.96    & 0.39   & 0.14  & 0.98   & 0.23  & 0.16     & 0.61  & 0.87   & 0.55    \\
503\_SWISSQUAL                             &  de & 54  & 4  & 24       & 0.93    & 0.33   & 0.23  & 0.93   & 0.34  & 0.18     & 0.65  & 0.92   & 0.57    \\
603\_TNO                                   &  nl & 48  & 4  & 24       & 0.94    & 0.33   & 0.23  & 0.97   & 0.26  & 0.16     & 0.60  & 0.84   & 0.67    \\
TUB\_DIS                                   &  de & 20  & 2  & 41       & 0.94    & 0.70   & 0.14  & 0.94   & 0.56  & 0.15     & 0.69  & 1.31   & 0.44    \\
TUB\_LIK                                   &  de & 8   & 12 & 20       & 0.99    & 0.22   & 0.11  & 0.99   & 0.28  & 0.16     & 0.69  & 0.84   & 1.05    \\
TUB\_VUPL                                  &  de & 15  & 4  & 36       & 0.89    & 0.76   & 0.21  & 0.70   & 0.63  & 0.26     & 0.93  & 1.10   & 0.18    \\
TUB\_PSA                                   &  en & 50  & 12 &  9       & 0.85    & 0.42   & 0.14  & 0.88   & 0.64  & 0.10     & 0.64  & 0.66   & 0.26    \\ \hline   
\textit{Average}                            &     &     &    &          & 0.92    & 0.45   & 0.19  & 0.91   & 0.43  & 0.19     & 0.60  & 0.97   & 0.57    \\ 
                                                                                                                                                      
\end{tabular}                                                                                                                                             
}
\end{table*}

The average and worst-case results of the proposed models across the 10 validation databases are shown in Table \ref{tab:results_1}. We also included two models (SM and SL) with the same architecture but without alignment and reference signal, as a single-ended baseline model. It should be noted that for each model the best result on average \textit{and} the best worst-case result is shown, which may be two different models. Interestingly the attention layer does not seem to help the model in improving the prediction accuracy (models with second letter L in Table \ref{tab:results_1}). This could be caused by the pretraining that is based on POLQA predictions. The training databases with subjective data may not be large enough to train the attention layer and POLQA itself may not perfectly model time dependencies. The models DM and DL that are using the dot product as similarity score outperform the single-ended models SM and SL, the performance difference is however small. Also, it can be noted that the accuracy of the single-ended model (RMSE* of 0.22) improved significantly when compared to the model in \cite{mittag2019c} (RMSE* of 0.29) without pretraining and end-to-end training. The LM model, which uses an L1-norm alignment with mean time pooling, obtains the best results and considerably improves the results when compared to the single-ended model.

\begin{figure}[!t]
\begin{minipage}[h]{1.0\linewidth}
\centering
\includegraphics[width=0.72\linewidth]{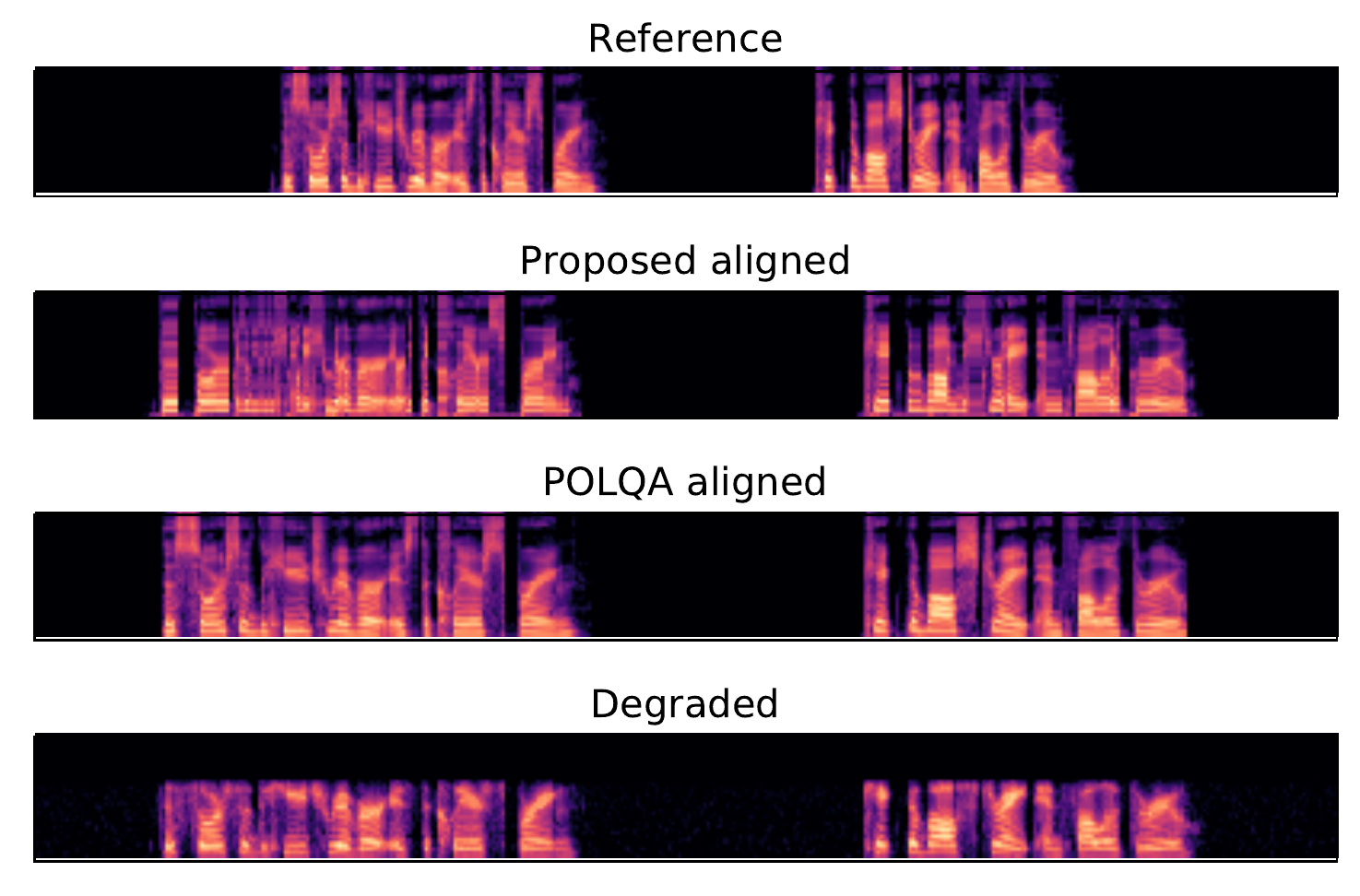}
\centerline{\small(a)}\medskip
\end{minipage}
%
%
\begin{minipage}[h]{1\linewidth}
\centering
\includegraphics[width=0.72\linewidth]{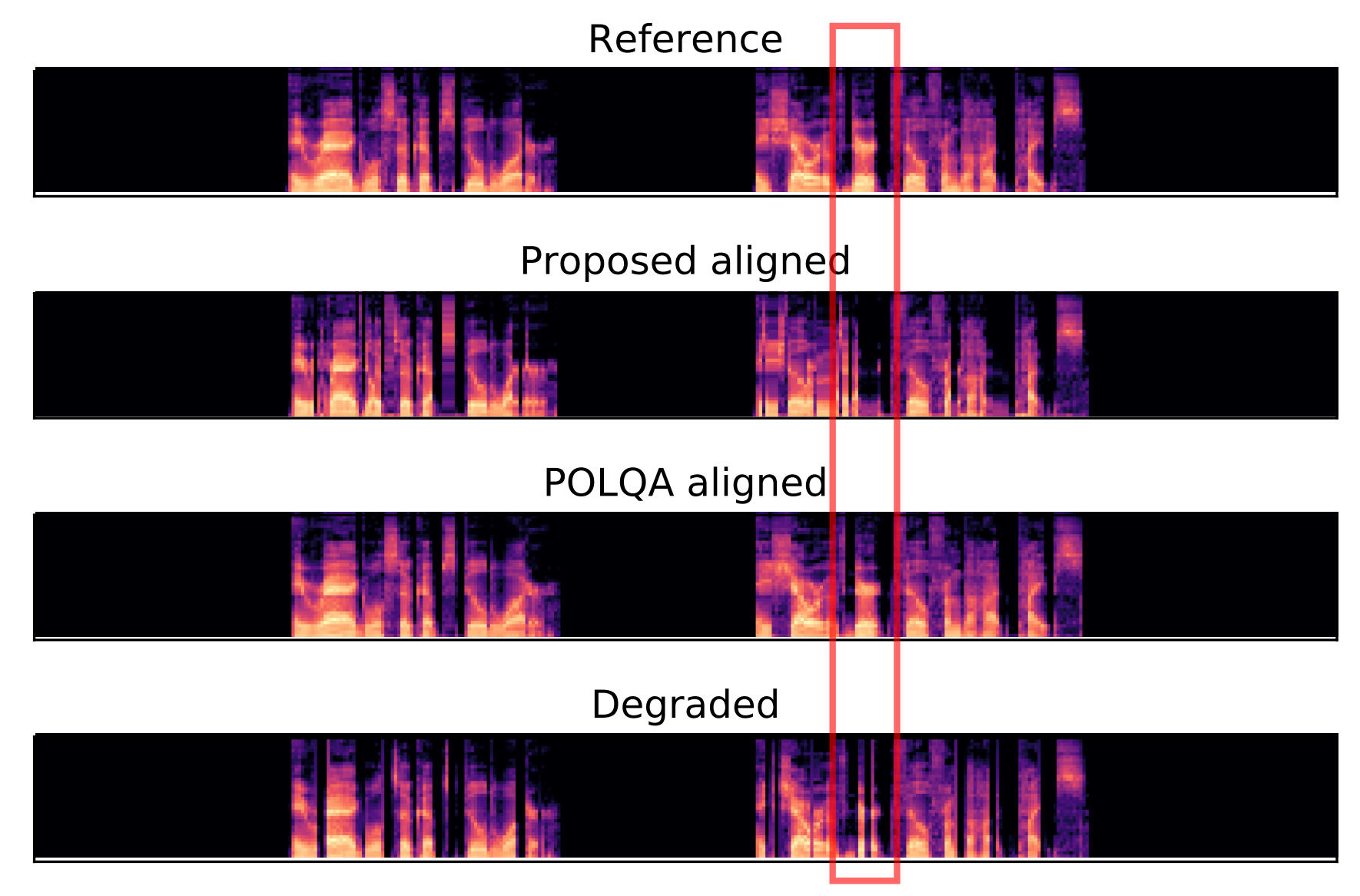}
\centerline{\small(b)} \medskip
\end{minipage}   
\caption{Time alignment of the proposed model compared to POLQA}
\label{fig:align}
\end{figure}

In Figure \ref{fig:align} a) the spectrogram of an unaligned degraded signal and its corresponding reference signal is shown. The time alignment of the proposed model with L1-norm alignment is compared to the time alignment by POLQA. It can be seen that the time alignment of the proposed model works well, especially considering that we did not implement any prior knowledge. For example, we don't restrict the model to look for future time steps only or time steps that are close to the current one. Figure \ref{fig:align} b) shows the spectrogram of a signal that is corrupted by time clipping but not exposed to delay. Inside the red box, it can be seen that some frames in the degraded signal were set to zero during a voiced segment. The proposed model aligns the wrong frames to these segments since it tries to find time steps that are similar to the clipped frames, whereas POLQA correctly aligns these time steps with the corresponding voiced segments of the reference speech signal.

Table \ref{tab:results_2} shows the results of the proposed model with L1-norm alignment and mean time pooling for all validation databases compared to POLQA and the full-reference model VISQOL \cite{Hines2015}. It can be seen that POLQA and the proposed model predict speech quality equally well with only marginal differences. This could mean that the model learned to copy the POLQA algorithm in the pretraining; however, the model accuracy after the pretraining and before the training with subjective MOS was significantly lower, somewhat contracting this theory. Still, it could be assumed that the model performance would further improve when more subjective data is available for training. Further, it should be noted that we selected the model that obtained the best results on the validation data set when trained on the training data set. Hence, the proposed model may be outperformed by POLQA on new, unseen test data. Regardless of this, the proposed architecture showed to be suitable for full-reference speech quality prediction with the advantage of being easily adaptable to new conditions, such as new codecs or noise suppression algorithms, through simple retraining of the model with subjective data.

\section{Conclusion}
\label{sec:con}
We presented a method to incorporate the clean reference speech signal into speech quality models that are based on deep learning. The approach uses a hard-attention mechanism to find, for each degraded time step, a corresponding time step in the reference feature set. The found aligned reference features can then be used to determine perceptual differences between both signals. Our experiments showed that the L1-norm difference of siamese network features is more suitable than using the dot product for the alignment. An attention layer for time pooling, on the contrary, does not seem to improve the prediction accuracy. We further showed that the prediction results improved by including the reference when compared to a single-ended baseline. The model achieved the same prediction accuracy as the current state-of-the-art model POLQA with an average RMSE* of 0.19. However, there is still room to improve the time alignment, for example, a step by step enrollment could be used, where time steps with longer distances are penalized in the loss function; or by training the network to find an optimal warping path \cite{NeuralWarp}.

\vfill\pagebreak
\bibliographystyle{IEEEbib}
\bibliography{refs}

\end{document}